\documentclass[useAMS,usenatbib]{mn2e}
\usepackage{multirow}
\usepackage{epsfig}
\usepackage{subfigure}
\usepackage{url}

\makeatletter
\let\jnl@style=\rm
\def\ref@jnl#1{{\jnl@style#1}}

\def\aj{\ref@jnl{AJ}}                   
\def\araa{\ref@jnl{ARA\&A}}             
\def\apj{\ref@jnl{ApJ}}                 
\def\apjl{\ref@jnl{ApJ}}                
\def\apjs{\ref@jnl{ApJS}}               
\def\ao{\ref@jnl{Appl.~Opt.}}           
\def\apss{\ref@jnl{Ap\&SS}}             
\def\aap{\ref@jnl{A\&A}}                
\def\aapr{\ref@jnl{A\&A~Rev.}}          
\def\aaps{\ref@jnl{A\&AS}}              
\def\azh{\ref@jnl{AZh}}                 
\def\baas{\ref@jnl{BAAS}}               
\def\jrasc{\ref@jnl{JRASC}}             
\def\memras{\ref@jnl{MmRAS}}            
\def\mnras{\ref@jnl{MNRAS}}             
\def\pra{\ref@jnl{Phys.~Rev.~A}}        
\def\prb{\ref@jnl{Phys.~Rev.~B}}        
\def\prc{\ref@jnl{Phys.~Rev.~C}}        
\def\prd{\ref@jnl{Phys.~Rev.~D}}        
\def\pre{\ref@jnl{Phys.~Rev.~E}}        
\def\prl{\ref@jnl{Phys.~Rev.~Lett.}}    
\def\pasp{\ref@jnl{PASP}}               
\def\pasj{\ref@jnl{PASJ}}               
\def\qjras{\ref@jnl{QJRAS}}             
\def\skytel{\ref@jnl{S\&T}}             
\def\solphys{\ref@jnl{Sol.~Phys.}}      
\def\sovast{\ref@jnl{Soviet~Ast.}}      
\def\ssr{\ref@jnl{Space~Sci.~Rev.}}     
\def\zap{\ref@jnl{ZAp}}                 
\def\nat{\ref@jnl{Nature}}              
\def\iaucirc{\ref@jnl{IAU~Circ.}}       
\def\aplett{\ref@jnl{Astrophys.~Lett.}} 
\def\apspr{\ref@jnl{Astrophys.~Space~Phys.~Res.}}
\def\bain{\ref@jnl{Bull.~Astron.~Inst.~Netherlands}}
\def\fcp{\ref@jnl{Fund.~Cosmic~Phys.}}  
\def\gca{\ref@jnl{Geochim.~Cosmochim.~Acta}}   
\def\grl{\ref@jnl{Geophys.~Res.~Lett.}} 
\def\jcp{\ref@jnl{J.~Chem.~Phys.}}      
\def\jgr{\ref@jnl{J.~Geophys.~Res.}}    
\def\jqsrt{\ref@jnl{J.~Quant.~Spec.~Radiat.~Transf.}}
\def\memsai{\ref@jnl{Mem.~Soc.~Astron.~Italiana}}
\def\nphysa{\ref@jnl{Nucl.~Phys.~A}}   
\def\physrep{\ref@jnl{Phys.~Rep.}}   
\def\physscr{\ref@jnl{Phys.~Scr}}   
\def\planss{\ref@jnl{Planet.~Space~Sci.}}   
\def\procspie{\ref@jnl{Proc.~SPIE}}   

\makeatother

\title[The X-ray reflector in NGC 4945]{The X-ray reflector in NGC 4945: a time and space resolved portrait}

\author[Andrea Marinucci, et al.]{A. Marinucci$^{1, 2}$\thanks{E-mail: marinucci@fis.uniroma3.it (AM)},  G. Risaliti$^{2,3}$,  Junfeng Wang$^2$,   E. Nardini$^2$, M. Elvis$^2$, G. Fabbiano$^2$, \newauthor S. Bianchi$^{1,}$,G. Matt$^1$ \\
$^1$Dipartimento di Fisica, Universit\`a degli Studi Roma Tre, via della Vasca Navale 84, 00146 Roma, Italy\\
$^2$Harvard-Smithsonian Center for Astrophysics, 60 Garden St., Cambridge MA 02138, USA\\
$^3$INAF - Osservatorio Astrofisico di Arcetri, L.go E. Fermi 5, Firenze, Italy\\
}

\begin{document}
\maketitle
\label{firstpage}

\begin{abstract} 
We present a time, spectral and imaging analysis of the X-ray reflector in NGC 4945, which reveals its geometrical and physical structure with unprecedented detail. NGC~4945 hosts one of the brightest AGN in the sky above 10~keV, but it is only visible through its reflected/scattered emission below 10~keV, due to absorption by a column density of $\sim$4$\times$10$^{24}$~cm$^{-2}$. 
A new {\em Suzaku} campaign of 5 observations spanning $\sim$6~months, together with past {\em XMM-Newton} and {\em Chandra} observations, show a remarkable constancy (within $<$10\%) of the reflected component. Instead, {\em Swift-BAT} reveals strong intrinsic variability on time scales longer than one year. Modeling the circumnuclear gas as a thin cylinder with the axis on the plane of the sky, we show that the reflector is at a distance $\geq$ 30-50~pc, well within the imaging capabilities of {\em Chandra} at the distance of NGC~4945 (1''$\sim$18~pc).
Accordingly, the {\em Chandra} imaging reveals a resolved, flattened, $\sim$150~pc-long clumpy structure, whose spectrum is fully due to cold reflection of the primary AGN emission. The clumpiness may explain the small covering factor derived from the spectral and variability properties.
\end{abstract}
\begin{keywords}
Galaxies: active - Galaxies: Seyfert - Galaxies: accretion
\end{keywords}
\section{ Introduction}
NGC 4945 is an almost edge-on (inclination angle $\sim80^{\circ}$) spiral galaxy hosting one of the nearest AGN (D $\sim$3.7~Mpc, 1''=18 pc, Mauersberger et al.~1996).
NGC 4945 has been widely studied in the past, both in the soft and in the hard X-rays. It is the brightest Seyfert 2 and
the second brightest radio-quiet AGN after NGC 4151 in the 100 keV sky (Done et al.~1996), with a strongly absorbed ($N_H\sim4\times 10^{24}$ cm$^{-2}$, Itoh et al.~2008) intrinsic nuclear continuum,  known to be extremely variable (Guainazzi et al.~2000, Madejski et al.~2000). Such a high column density completely blocks the primary nuclear
 emission below 8-10 keV while the emission at higher energy is still visible, though heavily affected by  Compton scattering and photoelectric absorption. 
A detailed analysis of the high energy variability (Madejski et al.~2000, Done et al.~2003, hereafter D03) already provided important (and, so far, almost unique) constraints on the geometry of the reflector: the  variability above 10~keV implies that most of the  observed emission is due to the primary emission, and not to Compton scattering from other directions (which would dilute the intrinsic variability, producing an almost constant observed emission). This, in turn, means that the reflector covers a solid angle $<$10~$\deg$ as seen from the source (assuming a toroidal structure).  This is also in agreement with the unusually low ratio between the observed, reflected emission below 10~keV, and the intrinsic flux in the same band as estimated from the emission above 10~keV ($f<$0.1\%). 
Finally, a {\em Chandra} observation revealed a spatially resolved component, on a $\sim$100~pc scale, whose spectrum is typical of Compton reflection from a $\sim$neutral medium (D03). 

Here we present the analysis of new and archival {\em Suzaku}, {\em XMM-Newton} and {\em Chandra} observations, focusing specifically on the determination, with unprecedented detail,  of the dimensions and the geometrical structure of the circumnuclear absorber/reflector in NGC~4945. A complete study, discussing both the {\em Chandra} imaging data, and our new {\em Suzaku} observations, will be presented in a separate paper (Marinucci et al.~2012, in prep.). The work presented here expands the results summarized above, through two completely new studies: \\
- the comparison between the intrinsic variability {\it above} 10~keV (from {\em Swift/BAT} observations), and the reflection variability {\it below} 10~keV (from a set of 8 {\em XMM-Newton} and {\em Suzaku} observations over a period of $\sim$10 years);\\
- a detailed analysis of the spatial extension of the reflector, taking advantage of the full set of available {\em Chandra} observations, as discussed in the next Sections.

\section{Observations, and data reduction}
\begin{figure}
  \includegraphics[width=8.3cm]{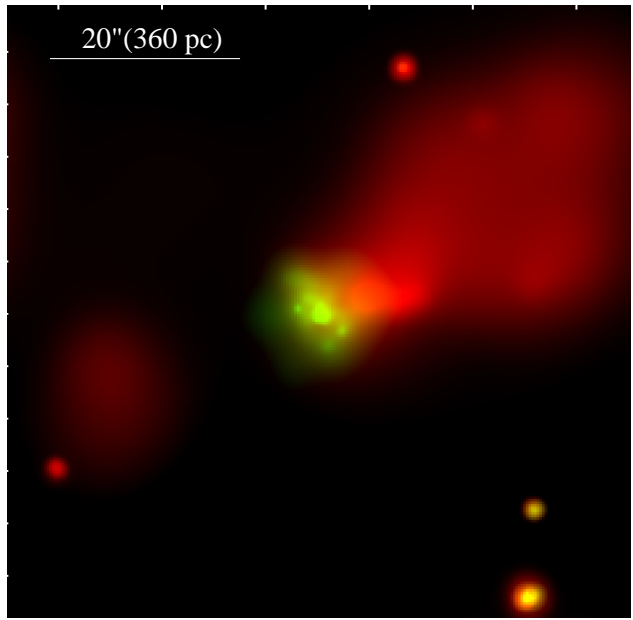}
  \caption{
Two-color (red:0.3-2~keV; green: 2-10~keV) {\em Chandra} image of the 1'$\times$1' central region of NGC~4945.
}
  \label{image}
\end{figure}

{\bf Chandra.} NGC 4945 was observed by {\em Chandra} on 2000, January 27-28 for a total
exposure time of 49 ks, with the ACIS camera. 
It was observed again four  years later twice,
for a total exposure time of 180~ks taking advantage of the HETG instrument.
Data were reduced with the CIAO~4.3 package (Fruscione et al.~2006)
and using the
Chandra Calibration Data Base (CALDB) version 4.4.6,
adopting standard procedures. 
The imaging analysis was performed on a file including the two merged HETG-zeroth order images reprojected on the ACIS-S image, applying the SER and smoothing procedures widely discussed in the literature (Tsunemi et al.~2001, Li et al.~2004, Wang et al.~2011), we therefore use a pixel size of 0.246 arcsec in Fig. \ref{image} and 0.123 arcsec in Fig. \ref{image2}. 
After cleaning for background flaring we get a total of 34 ks and 173 ks for the 2000 and the 2004 HETG-zeroth order merged observations, respectively.
Spectra were extracted on both the data set from three different regions: a circle with 25" radius (matching the {\em XMM-Newton} extraction region, see below); a $12''\times6''$ box, 
and a central circle with 1.5 arcsec radius. We grouped all extracted spectra to have at least 20 total counts per new bin.\\
{\bf Suzaku.} The six \textit{Suzaku} observations analysed in this paper were performed with the 
X-ray Imaging Spectrometer (XIS). The first one has been performed in 2006, has a duration of $\sim$100~ks, and has been presented by Itoh et al.~(2008). The remaining five observations, with a duration of 40~ks each, have been performed between July 2010 and January 2011, as part of a monitoring campaign of this source. 
  
The event files were processed and calibrated
adopting standard 
procedures, discussed in several previous papers (e.g. Maiolino et al.~2010).
The source extraction radius is $1.85'$ for all the 6 observations. Background spectra have been extracted from source-free regions with $3'$ radii.
The 0.5-10 keV spectra extracted from the front-illuminated XIS0 and XIS3 have been co-added.\\
{\bf XMM-Newton.} We used 2 XMM-\textit{Newton} observations performed with the EPIC-PN camera 
operated in large window and  medium filter modes. Source data 'cleaning' (exclusion of flaring particle background intervals) were performed  with SAS 10.0.0 (Gabriel et al.~2004) via an iterative process which leads to a maximization of the Signal-to-Noise Ratio (SNR), similarly to  that described in Piconcelli et al.~2004. Source spectra were extracted from a 25" radius circle; background spectra were extracted from source-free circular regions of the source field.\\
{\bf Swift.} The {\em Swift/BAT} 15-195~keV light curve has been obtained from the on-line catalog\footnote{http://swift.gsfc.nasa.gov/docs/swift/results/bs58mon/} which at the time of our analysis was available for 65 months. 
\section{The reflector geometry from time and spectral variability}
NGC~4945 offers a {\em unique} possibility to perform a comparison between the variability of the intrinsic and the reflected X-ray emission, because it is the only know AGN with both (a) Compton thickness $\tau_C$$>$1, and (b) a strong intrinsic variability above 10~keV. For {\em all} the other known bright AGN, such a detailed direct comparison between the variability of the two components is impossible because one of the two above mentioned conditions is not satisfied:\\
- (a) if the AGN is not Compton-thick, the Compton reflection component is diluted by the dominant direct component. The separation of the two components is never precise, and typical uncertainties in the flux of the reflected component are of the order of 20-30\% even for long observations of the brightest AGN.\\
- (b) the AGN is Compton-thick, but the intrinsic emission is either not visible (due to N$_H$$>$10$^{25}$~cm$^{-2}$, as in NGC~1068), or, anyway, not variable, probably because of the dilution due to multiple Compton scatterings (such as in the Circinus galaxy).\\
NGC~4945 shows one of the brightest reflection dominated spectra below 10~keV, with  flux measurement uncertainties of the order of $\sim$1\%, and a high-quality light curve of its intrinsic emission above 10~keV, provided by {\em Swift/BAT}. 
The highest quality spectra below 10~keV have been obtained in the eight {\em XMM-Newton} and {\em Suzaku} observations presented in the previous Section. 

The 3-10~keV spectra of these observations
have been fitted with a model consisting of a reflection continuum (PEXRAV, Madgziarz \& Zdziarski~1995), three emission lines for neutral, He-like and H-like iron, and an additional power law, requested by the fit, and associated to the diffuse emission (see below). From the analysis of the HETG grating spectra we rule out the presence of emission features associated to the Iron K$\beta$ line.
The analysis has been performed with the XSPEC~12.7 code (Arnaud et al.~1996). 
All quoted errors are at the 90\% confidence level for one interesting parameter. This model successfully reproduces all the individual spectra, with best fit parameters 
typical of reflection-dominated sources. The details of the spectral model have been discussed in Itoh et al.~2008, Schurch et al.~2002 and D03, and will be further analyzed in a forthcoming paper, together with the soft emission. What is interesting in this context is instead the possible  variability of the reflected component.
In this respect, the following results are relevant:\\
1) The analysis of the {\em Chandra} spectrum of the resolved 12$\times$6~arcsecond box, shows that the central emission is well fitted by a pure reflection continuum, with no need for an additional power law, so confirming that this component is needed only to reproduce the circumnuclear emission. Similarly, the {\em XMM-Newton} spectrum, extracted from the central 25", requires only a minor contribution from the additional power law (about 1/10 of that in the {\em Suzaku} spectra).\\
2) If we fit the eight spectra with constant reflection component and emission lines, leaving the additional power law as the only variable component, we obtain an equally good fit, with no significant residuals in any individual spectrum. This implies that formally, {\em all} the observed variability may be due to  galactic sources.\\
3) The whole luminosity within the central 1.85~arcmin region is only $\sim$2$\times10^{39}$~erg~s$^{-1}$. It is therefore not surprising if significant variability due to single sources is observed. \\
Based on the above considerations, we conclude that the best way to estimate the reflection variability is to fit the 8 spectra with the model described above, with only three free parameters: the slope and the normalization of the additional power law, and the flux F of the reflection (continuum plus lines) component, normalized to the value of the first interval, for ease of comparison.
The results are reported in Table~1, and show no significant variability in any observation. If we assume a constant value of the relative flux F, we obtain $<$F$>$=0.98$\pm$0.02, and a marginally better fit (the $\chi^2$ increases by 3, with 7 more degrees of freedom). The dispersion around the central value is significantly smaller  than the uncertainties on the individual values. This is due to the partial degeneracy among the spectral parameters. Based on these results, we estimate an upper limit to the observed reflection variability of $\sim$4\%, corresponding to twice the dispersion, or, equivalently, to the ratio between the maximum and minimum value of F within the measured dispersion.  
\begin{figure}
  \includegraphics[width=\columnwidth]{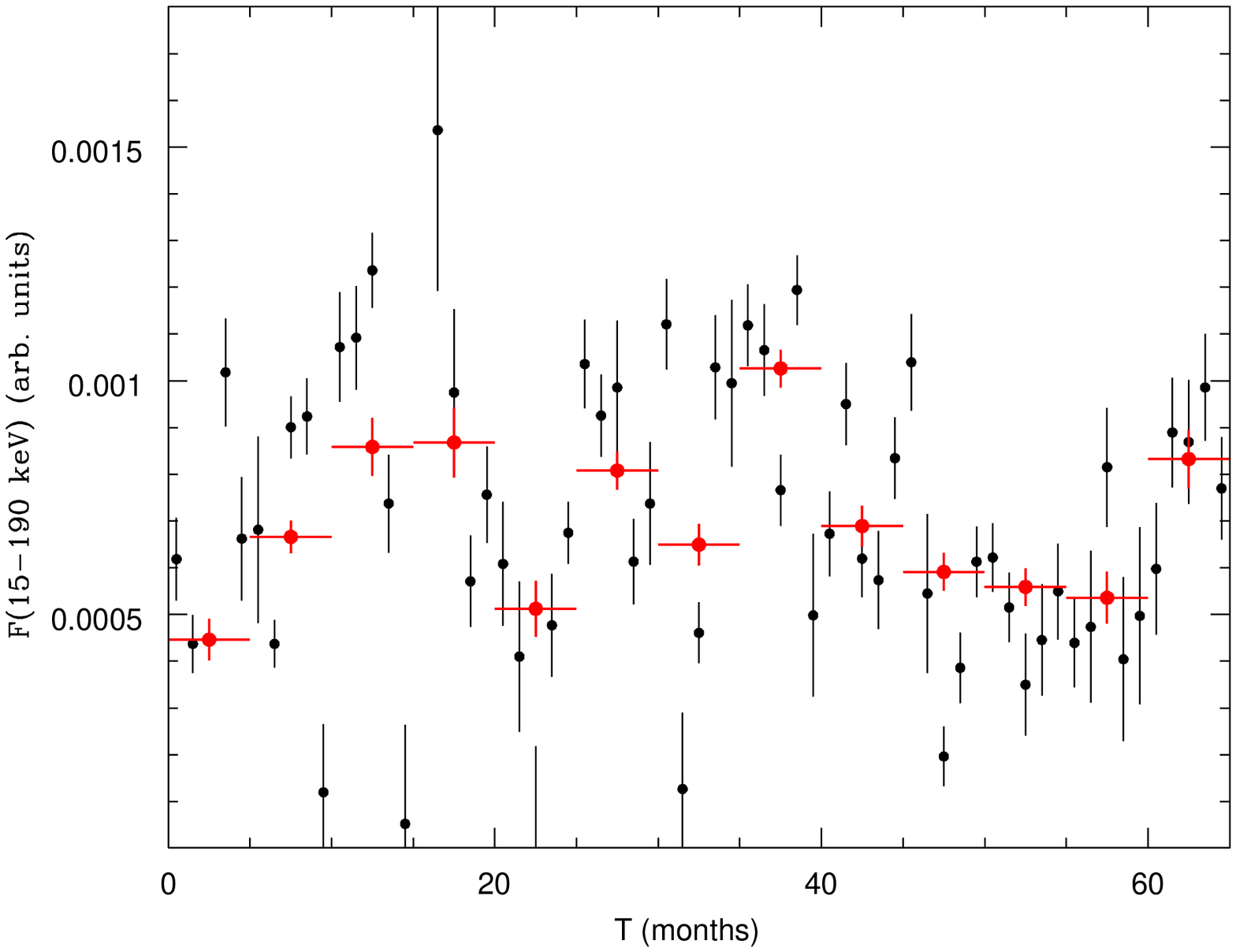}
  \includegraphics[width=\columnwidth]{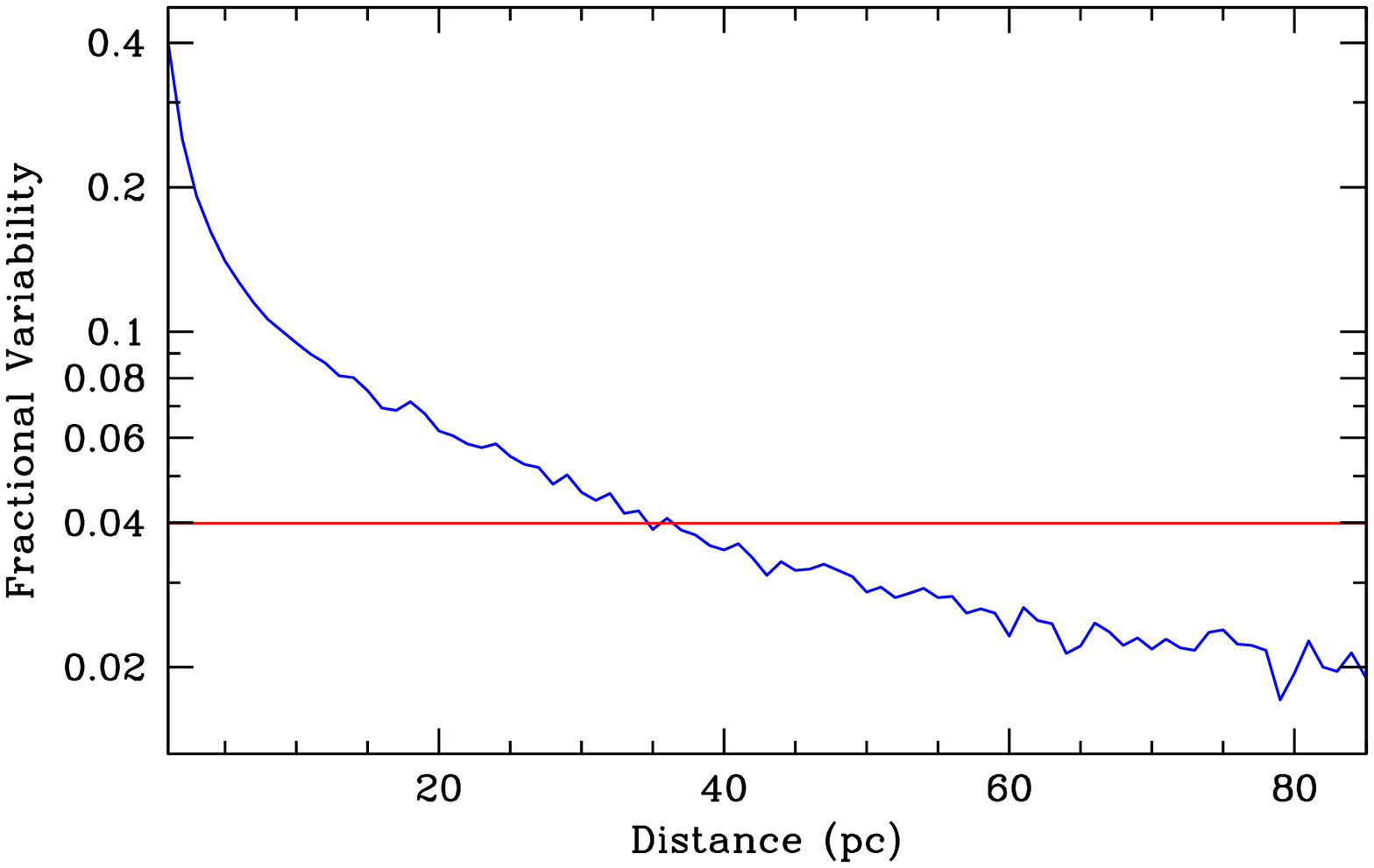}
  \caption{Upper panel: {\em Swift/BAT} light curve of the 15-195~keV emission of NGC~4945, grouped in time bins of 1~month (black) and five months (red). Lower panel: expected maximum variability as a response of a cylindrical reflector to the BAT light curve, as a function of the distance form the central source. The value corresponding to the upper limit to the observed reflection variability is $\sim$35~pc.}
  \label{bat}
\end{figure}
This result can be compared with the variability observed at higher energies. The average 15-150~keV flux of NGC~4945 is of the order of 5-10~mcrab, therefore the {\em Swift/BAT} monitoring provides reliable, high S/N light curves with reasonably short time bins. In Fig. \ref{bat} we show the 65-month light curve, in bins of one and five months. 
In order to obtain a semi-quantitative estimate of the reflector distance, we modeled the circumnuclear gas as a cylinder with the axis on the plane of the sky, with radius R and height H, such that H/R$<$0.1, as implied by the analysis of D03. We then used the observed BAT light curve as an input, and determined the maximum observable flux variation of the reflected component as a function of the distance R, by adding the contribution of each element of the reflector. Qualitatively, we expect that the more distant the reflector, the lower variability is observed in the reflected light curve, because the intrinsic variability is smoothed out by the different lengths of the light paths corresponding to each reflecting element. Quantitatively, the result is shown in Fig. \ref{bat}b. In order to observe a variation not larger than 4\% (reminding that $<$F$>$=0.96-1.00), the reflector must be at a distance larger than 35~pc. 
This result is obviously approximate, due to the assumptions on the geometry, and to the limited sampling of the reflection fluxes, which may have caught the source in the same reflected states by chance. However, we believe it provides a solid order of magnitude estimate of the distance of the reflector. We conclude noting that, considering that the {\em Chandra} resolution is $\sim0.25$~arcsec, ($\sim$5~pc at the distance of NGC~4945), the X-ray reflector should be easily resolved by {\em Chandra}, consistent with the conclusions of D03.
\begin{figure}
  \includegraphics[width=\columnwidth]{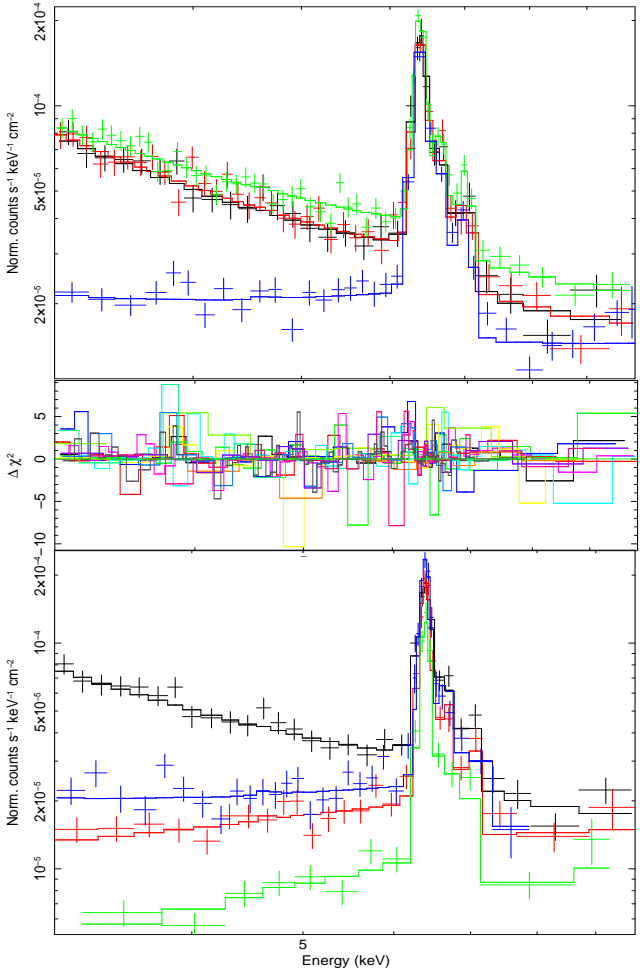}
  \caption{Spectra (rescaled to a common effective area), model and $\Delta\chi^2$ from the observations. The $\Delta\chi^2$ values are shown for all the spectra in the central panel, while only some of the spectra are shown, for clarity. Upper panel: first {\em Suzaku} observation in  2010 (black); {\em Suzaku} in 2011 (red), {\em Suzaku} in 2006 (green); {\em XMM-Newton} in 2004 (blue). Lower panel: the first 2010 {\em Suzaku} observation is shown again (black) for ease of comparison; the remaining spectra are from the {\em Chandra} 2004 data, extracted from a 25" radius (blue) a 12x6" box (red) and a 1.5" radius circle (green). It is worth noting that the spectra between 6 and 7~keV, dominated by the iron line emission, are almost identical in all observations, except for the last one from {\em Chandra}, and that the total continuum decreases with the area of the extraction region. Both these features are evidence that the variability is due to the diffuse galactic component, while the reflection remains constant, as quantitatively demonstrated by the spectral fits (Table~1). 
}
  \label{spectra}
\end{figure}
\section{The reflector geometry from direct imaging}
\begin{figure*}
  \includegraphics[width=16cm]{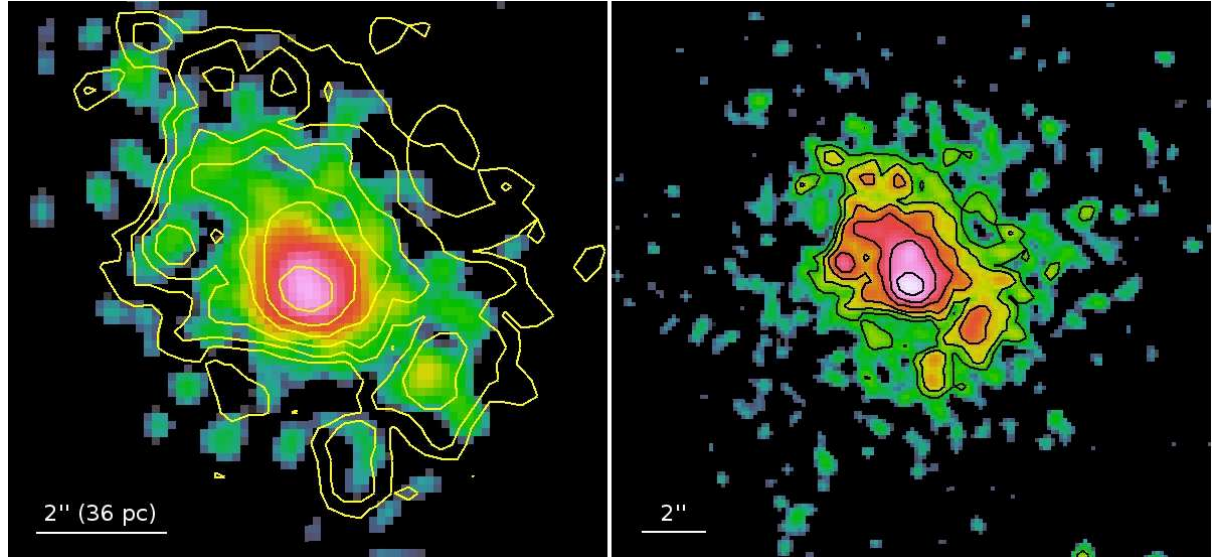}
  \caption{Image of the central region in the line emission band (6.2-6.7~keV, left) and in the hard X-ray band (right).}
  \label{image2}
\end{figure*}

In Fig. \ref{image} the 1'$\times$1' central region of NGC 4945 is shown. 
After running the script \textsc{chandra\_repro} on the three \textit{Chandra} event files we merged them all with the \textsc{merge\_all} script and filtered them in energy between 0.3-2 keV (in red) and 2-10 keV (in green).  A soft X-ray emitter can be seen in the top right region of the image, as already discussed in Schurch et al.~2002, where they called it the 'plume', reproduced by a single temperature thermal model (kT$\sim$ 0.6 keV), interpreted as a mass-loaded superwind emanated from the central nuclear starburst. With the inclusion of the two 2004 observations in the image, we notice a further soft lobe in the bottom-left direction indicating a second, possible emitter partially absorbed by a dust-lane crossing the nucleus, aligned with the clumpy structure shown in Fig. \ref{image2}. For this analysis we refer to a future work (Marinucci et  al., in prep.).\\
The nuclear 2-10 keV emission is embedded in a $12''\times6''$ box, corresponding to a 180 pc $\times$ 90 pc region, which is consistent with the starbust ring traced by molecular gas (Moorwood et al.~1996, Marconi et al.~2000, Curran et al.~2001).  

We performed a spectral analysis of three zones, corresponding to the {\em XMM-Newton} extraction region, a $12''\times6''$  box, including most of the hard emission (Fig. \ref{image}), and a central circular region with radius 1.5". The spectra, shown in Fig. \ref{spectra}, are fitted with the same reflection model as described above, with a contribution of the additional power law decreasing from the larger to the small region.
In particular: (a) the best fit model of the spectrum from the 25" circle is fully consistent (in each individual parameter) with that of the {\em XMM-Newton} observations; (b) the spectrum from the $12''\times6''$ box is again fitted with the same model, but with a relative flux of the reflection component F=0.86$\pm$0.07, indicating that most of the reflection comes from this region, with only a remaining $\sim$10\% from outer regions; (c) the model from the central annulus is fitted with a pure reflection component, with no need for the additional power law. The relative flux is F=0.58$\pm$0.06.
This shows that the reflection emission is moderately concentrated in the projected central $\sim30$~pc, but with a significant contribution from the outer regions. In our simple cylindrical scheme, if a small (a few degrees)  inclination angle is assumed, this concentration in the central unresolved region is expected due to the projection effect, and is not related to a small physical distance from the X-ray source.
A direct, complementary way to demonstrate that the hard X-ray emission is dominated by the reflected, diffuse component is by comparing the image in the full hard X-ray spectral interval (2-10~keV), with the exclusion of the 6.2-6.7 energy range, with the ``iron line image" obtained filtering only the 6.2-6.7~keV spectral interval (Fig. \ref{image2}), which is clearly dominated by the iron emission line (Fig. \ref{spectra}). We then overimposed the contours of the hard X-rays on the Iron K$\alpha$ emitter. The striking match between these two images demonstrates that the cold reflection and the iron emission lines originate exactly from the same circumnuclear material and we are actually able to take a screenshot of the inner reflecting structure of NGC~4945. 

Finally, we notice that the emission in Fig. \ref{image2} originates from a non homogeneous structure, with visible clumps and empty regions with sizes of the order of tens of parsecs.  The differences in counts among these sub-regions are highly significant (same area regions of 10$\times$10 pixels contain from $<1$0 to $>$100 counts). A smaller-scale clumpiness of these structures, not resolved by {\em Chandra}, may explain the low covering factor ($<$10\%) of the reflector, as estimated from the high intrinsic variability above 10~keV, and from the low ratio between the reflected and intrinsic components  (Madejski et al.~2000).

\begin{table}
\begin{center}
\begin{tabular}{lc|lcc}
\hline
Param. & Value & OBS & Date & F$^e$\\
\hline
$\Gamma$ & 1.6$\pm0.1$ & Suzaku 1 & 01/06 & 1.00$\pm$0.06\\
E(1)$^a$  & 6.39$\pm0.01$      & Suzaku 2 & 07/10 & 1.00$\pm$0.08\\           
EW(1)$^b$  & 1020$\pm100$ & Suzaku 3 & 07/10& 0.99$\pm$0.08\\
E(2)$^a$  &  6.66$\pm$0.01   & Suzaku 4 & 07/10& 0.97$\pm$0.08\\
EW(2)$^b$  & 130$\pm20$   & Suzaku 5 & 08/10& 0.98$\pm$0.08\\
E(3)$^a$  &  7.02$\pm$0.01   & Suzaku 6 & 01/11 &0.94$\pm$0.08\\
EW(3)$^b$  & 140$^{+20}_{-10}$  & XMM 1  & 01/00& 0.97$\pm$0.05\\
N$^c$      & 3.7$\pm0.7$ & XMM 2&01/04  & 0.98$\pm$0.06\\
F$_{2-10}^d$ & 1.3$\times$10$^{-12}$ & Chandra 25"&05/04 & 1.08$\pm$0.11\\
L$_{2-10}^d$ & 2.2$\times$10$^{39}$   & Chandra BOX&05/04 & 0.86$\pm$0.07\\
$\chi^2$/d.o.f.& 1308/1205  & Chandra 1.5"&05/04 &0.58$\pm$0.06\\
\hline
\end{tabular}\\
\caption{\label{bestfit} Best fit values for the reflection component. 
$^a$: emission line energies, in keV; $^b$: line equivalent widths, in eV; $^c$: normalization in units of 10$^{-3}$ ph cm$^{-2}$ s$^{-1}$ keV$^{-1}$; $^d$: 2-10~keV flux and luminosity of the reflection spectrum, in c.g.s. units. $^e$: relative flux.}
\end{center}
\end{table}

\section{Conclusions}
The time, spectral and imaging analysis of the X-ray emission of NGC~4945 based on {\em XMM-Newton}, {\em Suzaku}, {\em Swift/BAT} and {\em Chandra} data, characterizes of the AGN circumnuclear reflector with unprecedented detail.\\
1) The comparison between the strong intrinsic variability measured by {\em Swift/BAT} and the constant reflection spectra from {\em XMM-Newton} and {\em Suzaku} observations over more than ten years implies a distance of the reflector D$>$35~pc.\\
2) The {\em Chandra} image, obtained combining two observations performed in 2000 and 2004, reveals an extended hard emission on projected scales of $\sim$200$\times$100~pc. The spectrum of this emission is entirely reproduced by a cold reflection model. The central 30~pc account for about 50\% of the whole emission.\\
3) We show the first X-ray image of the inner reflector of an AGN. The large scale structure of the emission region is clumpy and asymmetric, and the central region is resolved by {\em Chandra}.   
 A smaller-scale, unresolved clumpiness may explain the low covering factor of the reflector inferred by the high variability of the intrinsic emission at E$>$10~keV, and by the low ratio between reflected and intrinsic spectrum (Madejski et al.~2000, D03).\\\\
\section*{Acknowledgements}
The authors are grateful to the referee for her/his useful comments. AM, GR and EM acknowledge NASA grants NNX11AC85G and NNX10AF50G. JW acknowledges support from NASA grants GO8-9101X and GO1-12009X.


\begin{thebibliography}{}

\bibitem[\protect\citeauthoryear{{Arnaud}}{{Arnaud}}{1996}]{xspec}
{Arnaud} K.~A.,  1996, in ASP Conf. Ser. 101: Astronomical Data Analysis
  Software and Systems V {XSPEC: The First Ten Years}.
p.~17

\bibitem[\protect\citeauthoryear{{Curran}, {Johansson}, {Bergman},
  {Heikkil{\"a}} \& {Aalto}}{{Curran} et~al.}{2001}]{curran01}
{Curran} S.~J.,  {Johansson} L.~E.~B.,  {Bergman} P.,  {Heikkil{\"a}} A.,
  {Aalto} S.,  2001, \aap, 367, 457

\bibitem[\protect\citeauthoryear{{Dickey} \& {Lockman}}{{Dickey} \&
  {Lockman}}{1990}]{dl90}
{Dickey} J.~M.,  {Lockman} F.~J.,  1990, \araa, 28, 215

\bibitem[\protect\citeauthoryear{{Done}, {Madejski} \& {Smith}}{{Done}
  et~al.}{1996}]{done96}
{Done} C.,  {Madejski} G.~M.,    {Smith} D.~A.,  1996, \apjl, 463, L63

\bibitem[\protect\citeauthoryear{{Done}, {Madejski}, {{\.Z}ycki} \&
  {Greenhill}}{{Done} et~al.}{2003}]{done03}
{Done} C.,  {Madejski} G.~M.,  {{\.Z}ycki} P.~T.,    {Greenhill} L.~J.,  2003,
  \apj, 588, 763

\bibitem[\protect\citeauthoryear{{Fruscione}, {McDowell}, {Allen},
  {Brickhouse}, {Burke}, {Davis}, {Durham}, {Elvis}, {Galle}, {Harris},
  {Huenemoerder}, {Houck}, {Ishibashi}, {Karovska}, {Nicastro}, {Noble},
  {Nowak} \& {Primini}}{{Fruscione} et~al.}{2006}]{ciao}
{Fruscione} A.,  {McDowell} J.~C.,  {Allen} G.~E.,  {Brickhouse} N.~S.,
  {Burke} D.~J.,  {Davis} J.~E.,  {Durham} N.,  {Elvis} M.,  {Galle} E.~C.,
  {Harris} D.~E.,  {Huenemoerder} D.~P.,  {Houck} J.~C.,  {Ishibashi} B.,
  {Karovska} M.,  {Nicastro} F.,  {Noble} M.~S.,  {Nowak} M.~A.,    {Primini}
  F.~A.,  2006, in Observatory Operations: Strategies, Processes, and Systems.
  Edited by Silva, David R.; Doxsey, Rodger E.. Proceedings of the SPIE, Volume
  6270, pp. 62701V (2006). Vol.~6270 of Presented at the Society of
  Photo-Optical Instrumentation Engineers (SPIE) Conference, {CIAO: Chandra's
  data analysis system}

\bibitem[\protect\citeauthoryear{{Gabriel}, {Denby}, {Fyfe}, {Hoar}, {Ibarra},
  {Ojero}, {Osborne}, {Saxton}, {Lammers} \& {Vacanti}}{{Gabriel}
  et~al.}{2004}]{gabr04}
{Gabriel} C.,  {Denby} M.,  {Fyfe} D.~J.,  {Hoar} J.,  {Ibarra} A.,  {Ojero}
  E.,  {Osborne} J.,  {Saxton} R.~D.,  {Lammers} U.,    {Vacanti} G.,  2004, in
  {F.~Ochsenbein, M.~G.~Allen, \& D.~Egret} ed., Astronomical Data Analysis
  Software and Systems (ADASS) XIII Vol.~314 of Astronomical Society of the
  Pacific Conference Series, {The XMM-Newton SAS - Distributed Development and
  Maintenance of a Large Science Analysis System: A Critical Analysis}.
pp 759--+

\bibitem[\protect\citeauthoryear{{Garmire}, {Bautz}, {Ford}, {Nousek} \&
  {Ricker}}{{Garmire} et~al.}{2003}]{acis}
{Garmire} G.~P.,  {Bautz} M.~W.,  {Ford} P.~G.,  {Nousek} J.~A.,    {Ricker}
  G.~R.,  2003, in X-Ray and Gamma-Ray Telescopes and Instruments for
  Astronomy. Edited by Joachim E. Truemper, Harvey D. Tananbaum. Proceedings of
  the SPIE, Volume 4851, p. 28-44 {Advanced CCD imaging spectrometer (ACIS)
  instrument on the Chandra X-ray Observatory}

\bibitem[\protect\citeauthoryear{{Greenhill}, {Moran} \&
  {Herrnstein}}{{Greenhill} et~al.}{1997}]{greenhill97}
{Greenhill} L.~J.,  {Moran} J.~M.,    {Herrnstein} J.~R.,  1997, \apjl, 481,
  L23

\bibitem[\protect\citeauthoryear{{Guainazzi}, {Matt}, {Brandt}, {Antonelli},
  {Barr} \& {Bassani}}{{Guainazzi} et~al.}{2000}]{gua00}
{Guainazzi} M.,  {Matt} G.,  {Brandt} W.~N.,  {Antonelli} L.~A.,  {Barr} P.,
  {Bassani} L.,  2000, \aap, 356, 463

\bibitem[\protect\citeauthoryear{{Heckman}, {Armus} \& {Miley}}{{Heckman}
  et~al.}{1990}]{heckman90}
{Heckman} T.~M.,  {Armus} L.,    {Miley} G.~K.,  1990, \apjs, 74, 833

\bibitem[\protect\citeauthoryear{{Itoh}, {Done}, {Makishima}, {Madejski},
  {Awaki}, {Gandhi}, {Isobe}, {Dewangan}, {Griffthis}, {Anabuki}, {Okajima},
  {Reeves}, {Takahashi}, {Ueda}, {Eguchi} \& {Yaqoob}}{{Itoh}
  et~al.}{2008}]{itoh08}
{Itoh} T.,  {Done} C.,  {Makishima} K.,  {Madejski} G.,  {Awaki} H.,  {Gandhi}
  P.,  {Isobe} N.,  {Dewangan} G.~C.,  {Griffthis} R.~E.,  {Anabuki} N.,
  {Okajima} T.,  {Reeves} J.~N.,  {Takahashi} T.,  {Ueda} Y.,  {Eguchi} S.,
  {Yaqoob} T.,  2008, \pasj, 60, 251

\bibitem[\protect\citeauthoryear{{Iwasawa}, {Koyama}, {Awaki}, {Kunieda},
  {Makishima}, {Tsuru}, {Ohashi} \& {Nakai}}{{Iwasawa}
  et~al.}{1993}]{iwasawa93}
{Iwasawa} K.,  {Koyama} K.,  {Awaki} H.,  {Kunieda} H.,  {Makishima} K.,
  {Tsuru} T.,  {Ohashi} T.,    {Nakai} N.,  1993, \apj, 409, 155

\bibitem[\protect\citeauthoryear{{Koornneef}}{{Koornneef}}{1993}]{koorn93}
{Koornneef} J.,  1993, \apj, 403, 581

\bibitem[\protect\citeauthoryear{{Li}, {Kastner}, {Prigozhin}, {Schulz},
  {Feigelson} \& {Getman}}{{Li} et~al.}{2004}]{li04}
{Li} J.,  {Kastner} J.~H.,  {Prigozhin} G.~Y.,  {Schulz} N.~S.,  {Feigelson}
  E.~D.,    {Getman} K.~V.,  2004, \apj, 610, 1204

\bibitem[\protect\citeauthoryear{{Madejski}, {{\.Z}ycki}, {Done}, {Valinia},
  {Blanco}, {Rothschild} \& {Turek}}{{Madejski} et~al.}{2000}]{madejski00}
{Madejski} G.,  {{\.Z}ycki} P.,  {Done} C.,  {Valinia} A.,  {Blanco} P.,
  {Rothschild} R.,    {Turek} B.,  2000, \apjl, 535, L87

\bibitem[\protect\citeauthoryear{{Marconi}, {Oliva}, {van der Werf},
  {Maiolino}, {Schreier}, {Macchetto} \& {Moorwood}}{{Marconi}
  et~al.}{2000}]{marconi00}
{Marconi} A.,  {Oliva} E.,  {van der Werf} P.~P.,  {Maiolino} R.,  {Schreier}
  E.~J.,  {Macchetto} F.,    {Moorwood} A.~F.~M.,  2000, \aap, 357, 24

\bibitem[\protect\citeauthoryear{{Matt}}{{Matt}}{2002}]{matt02}
{Matt} G.,  2002, \mnras, 337, 147

\bibitem[\protect\citeauthoryear{{Mauersberger}, {Henkel}, {Whiteoak}, {Chin}
  \& {Tieftrunk}}{{Mauersberger} et~al.}{1996}]{mauersberger96}
{Mauersberger} R.,  {Henkel} C.,  {Whiteoak} J.~B.,  {Chin} Y.-N.,
  {Tieftrunk} A.~R.,  1996, \aap, 309, 705

\bibitem[\protect\citeauthoryear{{Moorwood}, {van der Werf}, {Kotilainen},
  {Marconi} \& {Oliva}}{{Moorwood} et~al.}{1996}]{moor96}
{Moorwood} A.~F.~M.,  {van der Werf} P.~P.,  {Kotilainen} J.~K.,  {Marconi} A.,
     {Oliva} E.,  1996, \aap, 308, L1

\bibitem[\protect\citeauthoryear{{Nakai}}{{Nakai}}{1989}]{nakai89}
{Nakai} N.,  1989, \pasj, 41, 1107

\bibitem[\protect\citeauthoryear{{Piconcelli}, {Jimenez-Bail{\' o}n},
  {Guainazzi}, {Schartel}, {Rodr{\'{\i}}guez-Pascual} \& {Santos-Lle{\'
  o}}}{{Piconcelli} et~al.}{2004}]{pico04}
{Piconcelli} E.,  {Jimenez-Bail{\' o}n} E.,  {Guainazzi} M.,  {Schartel} N.,
  {Rodr{\'{\i}}guez-Pascual} P.~M.,    {Santos-Lle{\' o}} M.,  2004, \mnras,
  351, 161

\bibitem[\protect\citeauthoryear{{Rice}, {Lonsdale}, {Soifer}, {Neugebauer},
  {Kopan}, {Lloyd}, {de Jong} \& {Habing}}{{Rice} et~al.}{1988}]{rice88}
{Rice} W.,  {Lonsdale} C.~J.,  {Soifer} B.~T.,  {Neugebauer} G.,  {Kopan}
  E.~L.,  {Lloyd} L.~A.,  {de Jong} T.,    {Habing} H.~J.,  1988, \apjs, 68, 91

\bibitem[\protect\citeauthoryear{{Schurch}, {Roberts} \& {Warwick}}{{Schurch}
  et~al.}{2002}]{schurch02}
{Schurch} N.~J.,  {Roberts} T.~P.,    {Warwick} R.~S.,  2002, \mnras, 335, 241

\bibitem[\protect\citeauthoryear{{Str{\"u}der}, {Briel}, {Dennerl}, {Hartmann},
  {Kendziorra}, {Meidinger}, {Pfeffermann}, {Reppin}, {Aschenbach},
  {Bornemann}, {Br{\" a}uninger}, {Burkert} \& {Elender}}{{Str{\"u}der}
  et~al.}{2001}]{struder01}
{Str{\"u}der} L.,  {Briel} U.,  {Dennerl} K.,  {Hartmann} R.,  {Kendziorra} E.,
   {Meidinger} N.,  {Pfeffermann} E.,  {Reppin} C.,  {Aschenbach} B.,
  {Bornemann} W.,  {Br{\" a}uninger} H.,  {Burkert} W.,    {Elender} M.,  2001,
  \aap, 365, L18

\bibitem[\protect\citeauthoryear{{Tsunemi}, {Mori}, {Miyata}, {Baluta},
  {Burrows}, {Garmire} \& {Chartas}}{{Tsunemi} et~al.}{2001}]{tsunemi01}
{Tsunemi} H.,  {Mori} K.,  {Miyata} E.,  {Baluta} C.,  {Burrows} D.~N.,
  {Garmire} G.~P.,    {Chartas} G.,  2001, \apj, 554, 496

\bibitem[\protect\citeauthoryear{{Wang}, {Fabbiano}, {Risaliti}, {Elvis},
  {Karovska}, {Zezas}, {Mundell}, {Dumas} \& {Schinnerer}}{{Wang}
  et~al.}{2011}]{wang11}
{Wang} J.,  {Fabbiano} G.,  {Risaliti} G.,  {Elvis} M.,  {Karovska} M.,
  {Zezas} A.,  {Mundell} C.~G.,  {Dumas} G.,    {Schinnerer} E.,  2011, \apj,
  729, 75

\bibitem[\protect\citeauthoryear{{Whiteoak}, {Dahlem}, {Wielebinski} \&
  {Harnett}}{{Whiteoak} et~al.}{1990}]{whiteoak90}
{Whiteoak} J.~B.,  {Dahlem} M.,  {Wielebinski} R.,    {Harnett} J.~I.,  1990,
  \aap, 231, 25

\bibitem[\protect\citeauthoryear{{Wilson}, {Roy}, {Ulvestad}, {Colbert},
  {Weaver}, {Braatz}, {Henkel}, {Matsuoka}, {Xue}, {Iyomoto} \&
  {Okada}}{{Wilson} et~al.}{1998}]{wilson98}
{Wilson} A.~S.,  {Roy} A.~L.,  {Ulvestad} J.~S.,  {Colbert} E.~J.~M.,  {Weaver}
  K.~A.,  {Braatz} J.~A.,  {Henkel} C.,  {Matsuoka} M.,  {Xue} S.,  {Iyomoto}
  N.,    {Okada} K.,  1998, \apj, 505, 587

\end{thebibliography}


\begin{thebibliography}{}
\bibitem[\protect\citeauthoryear{Arnaud}{1996}]{1996ASPC..101...17A} Arnaud 
K.~A., 1996, Astronomical Data Analysis Software and Systems V, eds. Jacoby G. and Barnes J., p17, ASP Conf. Series volume 101.
\bibitem[\protect\citeauthoryear{Curran et al.}{2001}]{2001A&A...367..457C} Curran S.~J., Johansson L.~E.~B., Bergman P., Heikkil{\"a} A., Aalto S., 2001, A\&A, 367, 457
\bibitem[\protect\citeauthoryear{Done, Madejski,\& Smith}{1996}]{1996ApJ...463L..63D} Done C., Madejski G.~M., Smith D.~A., 1996, ApJ, 463, L63
\bibitem[\protect\citeauthoryear{Done et al.}{2003}]{2003ApJ...588..763D}
Done C., Madejski G.~M., {\.Z}ycki P.~T., Greenhill L.~J., 2003, ApJ, 588, 
763 (D03)
\bibitem[\protect\citeauthoryear{Fruscione et
al.}{2006}]{2006SPIE.6270E..60F} Fruscione A., et al., 2006, SPIE, 6270,
\bibitem[\protect\citeauthoryear{Gabriel et
al.}{2004}]{2004ASPC..314..759G} Gabriel C., et al., 2004, ASPC, 314, 759
\bibitem[\protect\citeauthoryear{Guainazzi et
al.}{2000}]{2000A&A...356..463G} Guainazzi M., Matt G., Brandt W.~N., Antonelli L.~A., Barr P., Ba
ssani L., 2000, A\&A, 356, 463
\bibitem[\protect\citeauthoryear{Itoh et al.}{2008}]{2008PASJ...60S.251I}
Itoh T., et al., 2008, PASJ, 60, 251
\bibitem[\protect\citeauthoryear{Li et al.}{2004}]{2004ApJ...610.1204L} Li
J., Kastner J.~H., Prigozhin G.~Y., Schulz N.~S., Feigelson E.~D., Getman
K.~V., 2004, ApJ, 610, 1204
\bibitem[\protect\citeauthoryear{Madejski et
al.}{2000}]{2000ApJ...535L..87M} Madejski G., {\.Z}ycki P., Done C.,Valinia A., Blanco P., Rothschild R., Turek B., 2000, ApJ, 535, L87
\bibitem[\protect\citeauthoryear{Magdziarz 
\& Zdziarski}{1995}]{1995MNRAS.273..837M} Magdziarz P., Zdziarski A.~A., 1995, MNRAS, 273, 837 
\bibitem[\protect\citeauthoryear{Maiolino et 
al.}{2010}]{2010A&A...517A..47M} Maiolino R., et al., 2010, A\&A, 517, A47 
\bibitem[\protect\citeauthoryear{Marconi et 
al.}{2000}]{2000A&A...357...24M} Marconi A., Oliva E., van der Werf P.~P., Maiolino R., Schreier E.~J., Macchetto F., Moorwood A.~F.~M., 2000, A\&A, 357, 24
\bibitem[\protect\citeauthoryear{Mauersberger et
al.}{1996}]{1996A&A...309..705M} Mauersberger R., Henkel C., Whiteoak J.~B., Chin Y.-N., Tieftrunk
 A.~R., 1996, A\&A, 309, 705 
\bibitem[\protect\citeauthoryear{Moorwood et 
al.}{1996}]{1996A&A...308L...1M} Moorwood A.~F.~M., van der Werf P.~P., Kotilainen J.~K., Marconi 
A., Oliva E., 1996, A\&A, 308, L1
\bibitem[\protect\citeauthoryear{Piconcelli et
al.}{2004}]{2004MNRAS.351..161P} Piconcelli E., Jimenez-Bail{\'o}n E.,
Guainazzi M., Schartel N., Rodr{\'{\i}}guez-Pascual P.~M., Santos-Lle{\'o}
M., 2004, MNRAS, 351, 161
\bibitem[\protect\citeauthoryear{Schurch, Roberts,
\& Warwick}{2002}]{2002MNRAS.335..241S} Schurch N.~J., Roberts T.~P., Warwick R.~S., 2002, MNRAS, 
335, 241 
\bibitem[\protect\citeauthoryear{Tsunemi et
al.}{2001}]{2001ApJ...554..496T} Tsunemi H., Mori K., Miyata E., Baluta C., 
Burrows D.~N., Garmire G.~P., Chartas G., 2001, ApJ, 554, 496 
\bibitem[\protect\citeauthoryear{Wang et al.}{2011}]{2011ApJ...742...23W}
Wang J., et al., 2011, ApJ, 742, 23 
\end{thebibliography}
\end{document}